\documentclass[12pt]{article}
\usepackage[centertags]{amsmath}
\usepackage{amssymb,amsfonts}
\usepackage{latexsym}
\usepackage{graphicx}
\begin{document}
\begin{center}
{\Large \bf Analytical models for gravitating radiating systems} \\
\vspace{1.5cm} {\bf B. P. Brassel\footnote{email: drbrasselint@gmail.com}, \bf S. D. Maharaj\footnote{email: maharaj@ukzn.ac.za
} and \bf G. Govender
}\\
Astrophysics and Cosmology Research Unit,\\
School of Mathematics, Statistics and Computer Science,\\
University of KwaZulu-Natal,\\
Private Bag X54001,\\
Durban 4000,\\
South Africa\\\vspace{1.5cm} {\bf Abstract}\\
\end{center}
We analyse the gravitational behaviour of a relativistic heat conducting fluid in a shear-free spherically symmetric spacetime. We show that the isotropy of pressure is a consistency condition which realises a second order nonlinear ordinary differential equation with variable coefficients in the gravitational potentials. Several new classes of solutions are found to the governing equation by imposing various forms on one of the potentials. Interestingly, a complex transformation leads to an exact solution with only real metric functions. All solutions are written in terms of elementary functions. We demonstrate graphically that the fluid pressure, energy density and heat flux are well behaved for the model, and the model is consistent with a core-envelope framework.


\section{Introduction}
Static spherically symmetric gravitational fields form the foundation for the description of highly dense objects in astrophysics where the matter distribution is normally considered to be a static perfect neutral or charged fluid. There exist, also, exact analytical solutions of the Einstein field equations for shear-free spacetimes, the earliest model being attributed to Kustaanheimo and Qvist \cite{kus&qvi}. A recent review of known solutions and generation techniques is provided by Ivanov \cite{ivan}. Shear-free models may also contain heat flow in the form of a non-vanishing radial heat flux through the interior and across the surface of a radiating star. Some models were obtained by Deng and Manheim \cite{den&man,den&manh} in the field of cosmology, and more recently by Govender and Thirukkanesh \cite{gov&thi} and Tewari \cite{mahs}, in an astrophysics context. The slightly older conformally flat radiating solutions were obtained by Banerjee \emph{et al} \cite{ban}, and were applied to relativistic radiating stars by Herrera \emph{et al} \cite{her,herr}, Govender \cite{mah&gov}, and Govender and Thirukkanesh \cite{gov&thi}. Stellar models in general, however, are continuums that have nonzero shear, acceleration and expansion; in this context, the Einstein field equations are highly nonlinear, coupled partial differential equations. In shearing spacetimes, very few solutions have been elucidated in the literature. The well established results are due to  Marklund and Bradley \cite{mar&bra} without heat flux and Naidu \emph{et al} \cite{nai} in radiating stars.

Exact solutions to Einstein's field equations that describe spherically symmetric manifolds with heat flow which form the foundation of a relativistic model in astrophysics and cosmology. Despite the existence of many classes of exact solutions, only a few of them are physically acceptable. There exist various methods for solving the Einstein field equations exactly; a geometric approach is the theory of Lie analysis of differential equations. The Lie analysis has been used by Wafo Soh and Mahomed \cite{mso} and Nyonyi \emph{et al} \cite{nyo, nyon} in analysing shear-free relativistic fluids in four and higher dimensions. Other techniques include the use of harmonic maps as well as numerical modus operandi. A comprehensive review of the methods and procedures utilized in the generation of solutions is provided by Stephani \emph{et al} \cite{ste}.

The shear-free assumption is often employed in the study of self gravitating spheres and studies of gravitational collapse. It has the added property that shear-free congruences undergoing homogeneous expansion are equivalent to the astrophysical homological conditions in the absence of dissipation as pointed out by Herrera \emph{et al} \cite{herre}. We should point out that the shear-free condition is unstable in the presence of dissipative fluxes. Herrera \emph{et al} \cite{herrer} considered the conditions of stability in the presence of dissipative fluxes for a geodesic fluid. Dissipative processes, local anisotropy of pressure and energy density clearly affect stability. However it was shown in \cite{herrer} that the shear-free congruences in addition, affects the propagation of time and stability.

It's possible to exactly solve the underlying differential equation, the condition of pressure isotropy, for shear-free fluids with heat flux without advanced mathematical methods. This is achieved by choosing one of the potentials and transforming the condition of pressure isotropy into a standard differential equation. The Einstein field equations for a shear-free model with heat conduction are generated, and we transform the pressure isotropy condition to a differential equation with variable coefficients. This governing equation is solved in its natural state by choosing multitudinous forms for the gravitational potentials. Several classes of new solutions are obtained in terms of elementary functions. We then perform a physical analysis for a particular solution arising from the consistency condition. We plot graphs for the gravitational potentials as well as graphs for the temporal and spatial evolutions of the matter variables.

\section{The model}
The metric for spherically symmetric spacetimes, in the absence of shearing stresses, can be written as
\begin{equation}
ds^2 = -A^2dt^2 + B^2[dr^2 + r^2(d\theta^2 + \sin^2\theta
d\phi^2)],\label{abmet}
\end{equation}
in isotropic and comoving coordinates $(x^a)=(t,r,\theta,\phi)$. The metric functions $A$ and $B$ depend on both the timelike coordinate $t$ and the radial coordinate $r$. The matter distribution is described by a relativistic fluid with energy momentum tensor
\begin{equation}
T^{ab} = (\rho+p)u^au^b+pg^{ab}+q^au^b+q^bu^a,\label{emten1}
\end{equation}
where $\rho$ is the energy density, $p$ is the isotropic (kinetic) pressure and $q^a$ is the heat flux vector $(q^au_a)=0$. These quantities are measured relative to a comoving fluid four-velocity \textit{{\bf u}} which is unit and timelike $(u^au_a=-1)$. The Einstein field equations $G_{ab}=T_{ab}$ can then be written as
\begin{subequations}
\label{0sheareinfield}
\begin{eqnarray}
\rho &=& \frac{3\dot{B}^2}{A^2B^2}-
\frac{1}{B^2}\left(\frac{2B^{\prime\prime}}{B} - \frac{B^{\prime
2}}{B^2} +
\frac{4B^{\prime}}{rB}\right),\label{0sheareinfield1}\\
%&&\nonumber\label{0sheareinfield1}\\
p &=& \frac{1}{A^2}\left(\frac{-2\ddot{B}}{B} -
\frac{\dot{B}^2}{B^2} +
\frac{2\dot{A} \dot{B}}{AB}\right) \nonumber\\
%&&\nonumber\\
& & + \frac{1}{B^2}\left(\frac{B^{\prime 2}}{B^2} +
\frac{2A^{\prime}B^{\prime}}{AB} +
\frac{2A^{\prime}}{rA} + \frac{2B^{\prime}}{rB}\right),\label{0sheareinfield2}\\
%&&\nonumber\\
p &=& -\frac{2\ddot{B}}{BA^2} + \frac{2\dot{A}\dot{B}}{BA^3} -
\frac{\dot{B}^2}{A^2B^2} + \frac{A^{\prime}}{rAB^2}\nonumber\\
%&&\nonumber\\
& & + \frac{B^{\prime}}{rB^3} + \frac{A^{\prime\prime}}{AB^2} -
\frac{B^{\prime 2}}{B^4} +
\frac{B^{\prime\prime}}{B^3},\label{0sheareinfield3}\\
%&&\nonumber\\
q &=& -\frac{2}{AB^2}\left(-\frac{\dot{B}^{\prime}}{B} +
\frac{B^{\prime}\dot{B}}{B^2} +
\frac{A^{\prime}}{A}\frac{\dot{B}}{B}\right).\label{0sheareinfield4}
\end{eqnarray}
\end{subequations}
The field equations $(\ref{0sheareinfield})$ are a system of highly nonlinear, coupled partial differential equations that describe the dynamics of the matter field with heat flux.

The fluid pressure is isotropic, and consequently equations (\ref{0sheareinfield2}) and (\ref{0sheareinfield3}) give rise to the consistency condition
\begin{eqnarray}
\frac{A_{rr}}{A}+\frac{B_{rr}}{B}=\left(2\frac{B_r}{B}
+\frac{1}{r}\right)\left(\frac{A_r}{A}+\frac{B_r}{B}\right).
\label{pisotropycon1}
\end{eqnarray}
This equation governs the gravitational behaviour of the radiating spacetime and needs to be be solved to produce an exact solution of the system $(\ref{0sheareinfield})$. Introducing the new variable
\begin{equation}
x=r^2,\nonumber
\end{equation}
the pressure isotropy condition $(\ref{pisotropycon1})$ can be written alternatively as
\begin{equation}
\left(\frac{1}{B}\right)A_{xx}+2A_x\left(\frac{1}{B}\right)_x-A\left(\frac{1}{B}\right)_{xx}=0,\label{modifiedform}
\end{equation}
where subscripts denote differentiation with respect to the new
variable $x$. Equation $(\ref{modifiedform})$ governs the behaviour of the radiating model.

\section{Exact solutions}
We now present three new solutions for the transformed condition of isotropic pressure $(\ref{modifiedform})$. These solutions are generated by making appropriate choices for one of the gravitational potentials and attempting to determine an integrable equation in terms of the other potential.
\subsection{Solution \textrm{I}: $B=\alpha x^{\beta n}$}
In an attempt to generate a new solution, we set
\begin{equation}
B(r,t)=\alpha x^{\beta n},
\end{equation}
in $(\ref{modifiedform})$ where $\alpha=\alpha(t)$, $\beta=\beta(t)$ and $n\in\mathbb{R}$. This reduces the pressure isotropy condition $(\ref{modifiedform})$ to
\begin{equation}
x^2A_{xx}-\left(2\beta n\right)xA_{x}-\beta n\left(\beta n+1\right)A=0,\label{A1}
\end{equation}
which is a second order Cauchy-Euler differential equation in $A$. Utilising the standard transformation $A=x^m$, where $m=m(t)$, generates the corresponding characteristic equation
\begin{equation}
m^2-(2\beta n+1)m-(\beta^2n^2+\beta n)=0,\nonumber
\end{equation}
with roots
\begin{equation}
m=\frac{1}{2}(2\beta n+1)\pm\sqrt{8\beta^2n^2+8\beta n+1}.\nonumber
\end{equation}
Finally, the general solution to $(\ref{A1})$ may be written as
\begin{eqnarray}
A(x,t)&=& \tau(t)x^{\frac{1}{2}(2\beta n+1)+\sqrt{8\beta^2n^2+8\beta n+1}}\nonumber\\
& & +\chi(t)x^{\frac{1}{2}(2\beta n+1)-\sqrt{8\beta^2n^2+8\beta n+1}},
\label{A12}
\end{eqnarray}
where $\tau(t)$ and $\chi(t)$ are functions of integration. Thus, the metric has the form
\begin{eqnarray}
ds^2 &=& -\left[\tau(t)r^{2\beta n+1+\sqrt{8\beta^2n^2+8\beta n+1}}\right.\nonumber\\
& & \left.+\chi(t)r^{2\beta n+1-\sqrt{8\beta^2n^2+8\beta n+1}}\right]^2dt^2\nonumber\\
& & +(\alpha r^{\beta n})^2\left[dr^2+r^2(d\theta^2+\sin^2\theta d\phi^2)\right].\label{A2}
\end{eqnarray}
This is a new category of exact solutions for a shear-free fluid exhibiting heat conduction where $\alpha$, $\beta$, $\tau$ and $\chi$ are all free temporal functions, and $n\in\mathbb{R}$.

\subsection{Solution \textrm{II}: $B^{-1}=\alpha k^{\beta x+\gamma}$}
We make an exponential choice for $\frac{1}{B}$ in (\ref{modifiedform}), so that
\begin{equation}
\frac{1}{B}=\alpha k^{\beta x+\gamma},
\end{equation}
where $\alpha=\alpha(t)$, $\beta=\beta(t)$, $\gamma=\gamma(t)$ and $k\in\mathbb{R}$. Then equation $(\ref{modifiedform})$ reduces to
\begin{equation}
A_{xx}+2\beta(\ln k)A_{x}-\beta^2(\ln k)^2A=0,\label{char1}
\end{equation}
which is a second order ordinary differential equation with constant coefficients. The characteristic equation of $(\ref{char1})$ is
\begin{equation}
m^2+2(\ln k)\beta m -(\ln k)^2 \beta^2=0,\nonumber
\end{equation}
and its roots are
\begin{equation}
m=-\beta(\ln k) \pm\sqrt{2}\beta(\ln k),\nonumber
\end{equation}
which are real and distinct. Hence the general solution to $(\ref{char1})$ is given by
\begin{equation}
A(x,t)=\nu(t)k^{\beta(t)[-1-\sqrt{2}]x}+\kappa(t)k^{\beta(t)[-1+\sqrt{2}]x},
\end{equation}
where $\nu(t)$ and $\kappa(t)$ are integration functions. The line element for this class of exact models is given by
\begin{eqnarray}
ds^2 &=& -\left[\nu(t)k^{\beta(t)[-1-\sqrt{2}]r^2}+\kappa(t)k^{\beta(t)[-1+\sqrt{2}]r^2}\right]^2dt^2\nonumber\\
& & + \alpha(t)^{-2}k^{-2[\beta(t)r^2+\gamma(t)]}\left[dr^2+r^2(d\theta^2+\sin^2\theta d\phi^2)\right].\label{feynas}
\end{eqnarray}
This class of exact solutions has a very simple form.

\subsection{Solution \textrm{IV}: $B^{-1}=\alpha(t)(\beta(t) x+A)$}
In an attempt to generate another class of exact solutions we seek another coupling of the gravitational potentials in (\ref{modifiedform}). In particular we set
\begin{equation}
\frac{1}{B}=\alpha(\beta x+A),
\end{equation}
where $\alpha=\alpha(t)$ and $\beta=\beta(t)$. This assumption reduces the master equation $(\ref{modifiedform})$ to
\begin{equation}
\beta xA_{xx}+2\beta A_{x}+2A_{x}^2=0,\label{vixen}
\end{equation}
which is nonlinear. In order to solve this equation we must reduce the order and we set
\begin{equation}
A_{x}=v,\nonumber
\end{equation}
where $v=v(x)$. Equation $(\ref{vixen})$ then becomes
\begin{equation}
\beta x\frac{dv}{dx}+2\beta v+2v^2=0,
\end{equation}
which is a separable equation. Integrating this equation yields
\begin{equation}
v=\frac{dA}{dx}=-\frac{\beta D}{D-x^2},
\end{equation}
where $D=D(t)$ arises in the integration. This equation is also separable and so can be integrated to yield the general solution
\begin{equation}
A(x,t)=-\chi(t)\tanh^{-1}\left(\frac{x}{\chi(t)}\right)+\eta(t),
\end{equation}
where $\chi(t)$ and $\eta(t)$ are integration functions. This expression, along with the assumption $B=(1/\alpha)(\beta x+A)^{-1}$ constitutes another solution to the consistency condition $(\ref{modifiedform})$. Expressing these in terms of the original variables $r$ and $t$, we get
\begin{subequations}
\label{tsolu3}
\begin{eqnarray}
A(r,t) &=& \eta(t)-\chi(t)\tanh^{-1}\left(\frac{r^2}{\chi(t)}\right),\label{tsolu31}\\
B(r,t) &=& \alpha(t)^{-1}[\beta(t) r^2+A(r,t)]^{-1},\label{tsolu32}
\end{eqnarray}
\end{subequations}
which is a further new solution in exact form to the field equations. The line element $(\ref{abmet})$ is then
\begin{eqnarray}
ds^2 &=& -\left[\eta(t)-\chi(t)\tanh^{-1}\left(\frac{r^2}{\chi(t)}\right)\right]^2dt^2\nonumber\\
& & + \alpha(t)^{-2}\left[\beta(t) r^2+\eta(t)-\chi(t)\tanh^{-1}\left(\frac{r^2}{\chi(t)}\right)\right]^{-2}\nonumber\\
& & \times[dr^2+r^2(d\theta^2+\sin^2\theta d\phi^2)].
\end{eqnarray}
This new exact solution of the Einstein field equations is also expressible in terms of elementary functions.

\section{A useful transformation}
Other choices of the gravitational potential $B$ may lead to new solutions of the differential equation (\ref{modifiedform}). However, it is not clear how to achieve this in a systematic way. Here we present another new class of exact solutions to (\ref{modifiedform}). This class of exact solutions has the interesting feature that it arises via a complex transformation. This approach may be useful in producing new solutions or insights into other equations of physical interest.

A rational functional choice can be made for $\frac{1}{B}$ in (\ref{modifiedform}), namely we choose
\begin{equation}
\frac{1}{B}=\frac{\alpha x^2}{\beta x+1},\nonumber
\end{equation}
where $\alpha=\alpha(t)$ and $\beta=\beta(t)$. Equation $(\ref{modifiedform})$ then reduces to
\begin{equation}
x^2(\beta x+1)^2A_{xx}+2x(\beta x+1)(\beta x+2)A_{x}-2A=0,
\label{rat1}
\end{equation}
which is a second order linear ordinary differential equation with variable coefficients. Differential equations of the form of equation $(\ref{rat1})$ can be found in the book by Polyanin and Zaitsev \cite{pol&zai}. Unfortunately, with such equations, transformations that reduce (\ref{rat1}) to standard form are not unique or obvious. Thus, a trial and error approach has to be employed to reduce (\ref{rat1}) to a simpler form.

In our case, we can proceed by setting
\begin{equation}
\Omega=i\sqrt{2}\left[\ln(x)-\ln(\beta x+1)\right],\label{jhorner}
\end{equation}
where $i=\sqrt{-1}$. The appearance of the complex quantity $i$ should not be troublesome as we eventually will obtain a real solution. It is in fact due to certain special qualities of the complex quantity, i.e. the fact that its square is real and that its multiplicative inverse is its additive inverse, that allows for this. Then, equation $(\ref{rat1})$ becomes
\begin{equation}
\frac{\left(\frac{1}{\beta e^{i\Omega/\sqrt{2}}-1}+1\right)^2}{(\beta e^{i\Omega/\sqrt{2}}-1)^2}A_{xx}+2\frac{\left(\frac{1}{\beta e^{i\Omega/\sqrt{2}}-1}+1\right)\left(\frac{1}{\beta e^{i\Omega/\sqrt{2}}-1}+2\right)}{\beta e^{i\Omega/\sqrt{2}}-1}A_{x}-2A=0.\label{rat2}
\end{equation}
Noting that with the expressions $A_{x}=\frac{d\Omega}{dx}A_{\Omega}$ and $A_{xx}=\left(\frac{d\Omega}{dx}\right)^2A_{\Omega\Omega}+\frac{d^2\Omega}{dx^2}A_{\Omega}$,
equation $(\ref{rat2})$ reduces, after some calculation, to
\begin{equation}
-2A_{\Omega\Omega}+3i\sqrt{2}A_{\Omega}-2A=0,\label{trat}
\end{equation}
which is a second order linear differential equation with constant coefficients. Its characteristic equation is given by
\begin{equation}
2m^2-3i\sqrt{2}m+2=0,\nonumber
\end{equation}
which has roots
\begin{equation}
m=\frac{3i}{2\sqrt{2}}\pm\frac{1}{2}i\sqrt{\frac{17}{2}}.\nonumber
\end{equation}
Hence, the general solution to $(\ref{trat})$ is given by
\begin{equation}
A=\gamma e^{\left(\frac{3i}{2\sqrt{2}}+\frac{1}{2}i\sqrt{\frac{17}{2}}\right)\Omega}+\varphi e^{\left(\frac{3i}{2\sqrt{2}}-\frac{1}{2}i\sqrt{\frac{17}{2}}\right)\Omega}.\nonumber
\end{equation}
When we substitute for $\Omega$ from (\ref{jhorner}) we find that
\begin{eqnarray}
A(x,t) &=& \gamma(t)\left(\frac{x}{\beta(t) x+1}\right)^{-\frac{1}{2}\left(3+\sqrt{17}\right)}+\varphi(t)\left(\frac{x}{\beta(t) x+1}\right)^{\frac{1}{2}\left(-3+\sqrt{17}\right)},\label{ratsol}
\end{eqnarray}
where $\gamma(t)$ and $\varphi(t)$ are integration functions. Observe that the solution is given in real functions only. In terms of the original variables we can write
\begin{subequations}
\label{tsolu6}
\begin{eqnarray}
A(r,t) &=& \gamma(t)\left(\frac{r^2}{\beta(t) r^2+1}\right)^{-\frac{1}{2}\left(3+\sqrt{17}\right)}\nonumber\\
& & +\varphi(t)\left(\frac{r^2}{\beta(t) r^2+1}\right)^{\frac{1}{2}\left(-3+\sqrt{17}\right)},\label{tsolu61}\\
B(r,t) &=& \frac{\beta(t) r^2+1}{\alpha(t) r^4}.\label{tsolu62}
\end{eqnarray}
\end{subequations}
We have verified that (\ref{tsolu6}) satisfies the differential equation (\ref{rat1}) with the help of MATHEMATICA which gives the potentials. Thus the metric line element $(\ref{abmet})$ becomes
\begin{eqnarray}
ds^2 &=& -\left[\gamma(t)\left(\frac{r^2}{\beta(t) r^2+1}\right)^{-\frac{1}{2}\left(3+\sqrt{17}\right)}+\varphi(t)\left(\frac{r^2}{\beta(t) r^2+1}\right)^{\frac{1}{2}\left(-3+\sqrt{17}\right)}\right]^2dt^2\nonumber\\
& & + \left(\frac{\beta(t) r^2+1}{\alpha(t) r^4}\right)^{2}\left[dr^2+r^2(d\theta^2+\sin^2\theta d\phi^2)\right].
\end{eqnarray}
We have generated another class of exact solutions to the Einstein field equations which we believe is new and has not been found before. This class is made possible because (\ref{rat1}) can be transformed via a complex transformation to (\ref{trat}) with constant coefficients. The metric functions turn out to be expressible in terms of \textit{real functions} only. This is an unusual feature, and it is interesting to see this application arising in spherically symmetric gravitational fields. It is possible that the approach applied in this section may be used in other similar equations.

\section{Physical analysis}
In order for relativistic radiating solutions to be used as a basis in constructing physical models in astrophysics and cosmology, we need to first test the physics that is realised by a given exact solution. A comprehensive physical analysis is necessary to demonstrate the physical viability and applicability of a given model. However, due to the often complicated structure and nature of the solutions that are generated, a complete physical analysis is not always possible. A particular exact solution usually contains free parameters or functions that have to be fine tuned in order to generate good behaviour. We carry out a brief physical analysis for one of the exact solutions found in this paper.

Figures 1 and 2 feature spatial plots for the potentials of $(\ref{A2})$ at differing time slices. Both plots indicate smooth and monotonically increasing behaviour and in the case of $B$, the profile is almost constant. Despite the variations in the gravitational field seeming small, the field is still strong enough to have a marked influence on the matter. Next, we consider the matter quantities. We observe from Figure 3 that the energy density $\rho$ is smooth and regular throughout the interior. In the region close to the centre $r=0$ it is apparent that $\rho$ diverges. This solution is therefore valid in the outer regions of core-envelope models that are regarded as being more realistic for stellar interiors (for an example of a core envelope structure see the model of Paul and Tikekar \cite{pau&tik}). In Figure 4 we observe that the pressure $p$ from the solution is smooth and regular throughout the interior and  diverges at points closer to the centre of the sphere, as was the case with the energy density. It is interesting to note from Figures 3 and 4, that the energy density profile is noticeably steeper than the pressure profile, i.e., $d\rho/dt>dp/dt$ despite the fact that individual pressure changes are greater than the density changes (by two orders of magnitude), i.e., $\Delta p =(1.0\times 10^2)\Delta \rho$. Figure 5 indicates that the heat flux $q$ is smooth and monotonically decreasing from the centre outwards with divergence at the centre.
It is evident from Figure 5 that the profile of the heat flux is similar to that of the density. We also make the observation that $dq/dt > dp/dt$. Furthermore, Figures 3-5 reveal that $\Delta q>\Delta p>\Delta \rho$ and suggests that closer to the centre, $q>p>\rho$.

\newpage
\section{Discussion}
In this paper we have studied spherically symmetric shear-free spacetimes and the associated models that may be used to describe the interior of fluid spheres. We have generated several classes of new exact solutions to the field equations. Our models have vanishing shear and describe heat flow in the interior of the fluid spheres. The consistency condition arising from the isotropy of the fluid pressure was analysed and solved due to it being transformed into a standard differential equation, and by applying a complex transformation. In this investigation, particular forms for the gravitational potentials were chosen and several new solutions were obtained in terms of elementary functions. A physical analysis was performed, for a special case of the class of solutions with a power law form. The spatial and temporal profiles that were produced for the matter and gravitational variables, indicate that the resulting model is consistent with that of a core-envelope scenario, which is plausible for a realistic description of a stellar interior. Our results also suggest that closer to the centre of the fluid sphere, the heat flux $q$ dominates the pressure $p$ and energy density $\rho$ by at least two orders of magnitude. The investigations in this paper and the results generated form an essential part of a wide array of models that can be used within the framework of general relativity to construct realistic and physically meaningful studies in astrophysics and cosmology. For the purpose of astrophysical modeling, these investigations can be extended by including the features such as an equation of state, a finite boundary that localises the fluid distribution and acts as an interface between the interior and exterior spacetimes, and the dynamical stability of the dissipating fluid in the context of non-adiabatic gravitational collapse.

\begin{figure}[t]
\centering
\includegraphics[scale=1.5]{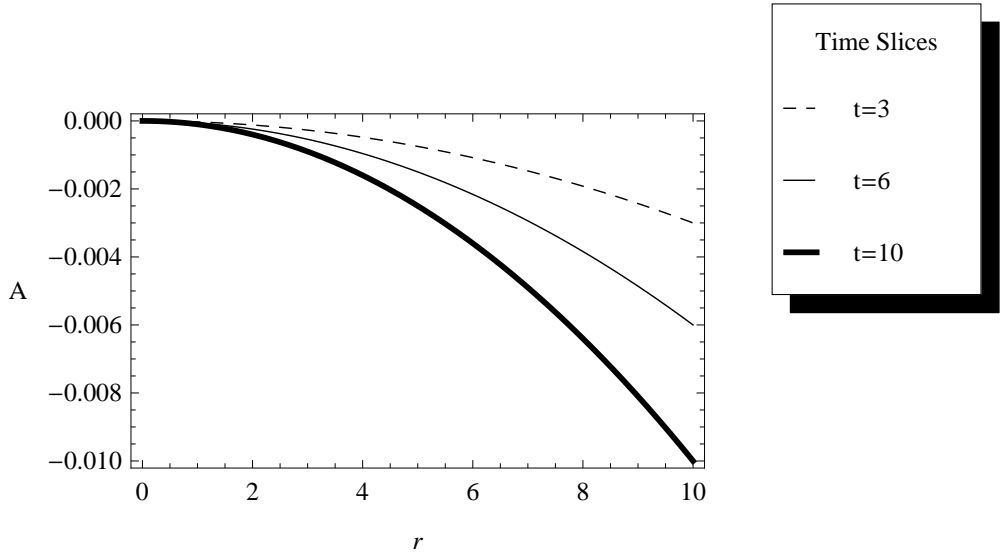} \caption{The potential $A$ versus $r$ at varying units of time.} \label{fig6}
\end{figure}

\begin{figure}[t]
\centering
\includegraphics[scale=1.5]{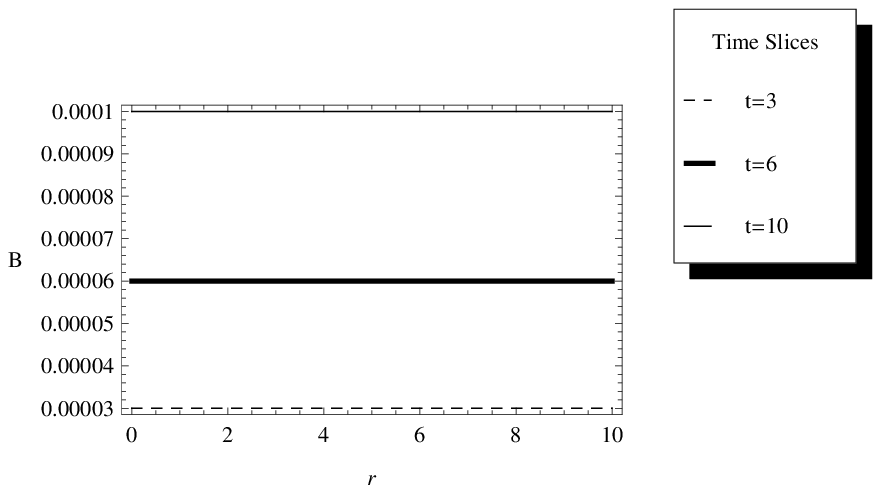} \caption{The potential $B$ versus $r$ at varying units of time.} \label{fig6}
\end{figure}

\begin{figure}[t]
\centering
\includegraphics[scale=1.5]{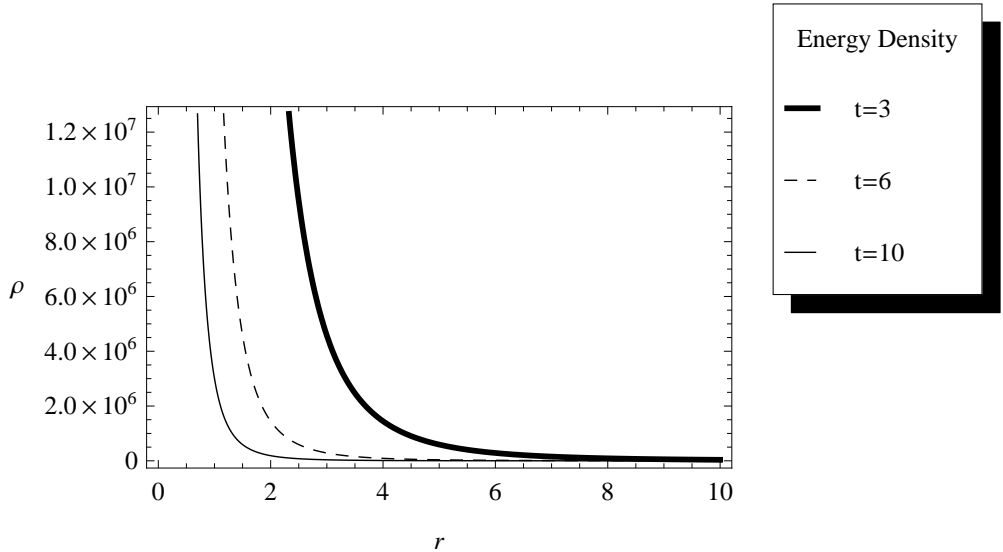} \caption{The radial profile for the energy density $\rho$ at various units of time.} \label{fig6}
\end{figure}

\begin{figure}[t]
\centering
\includegraphics[scale=1.5]{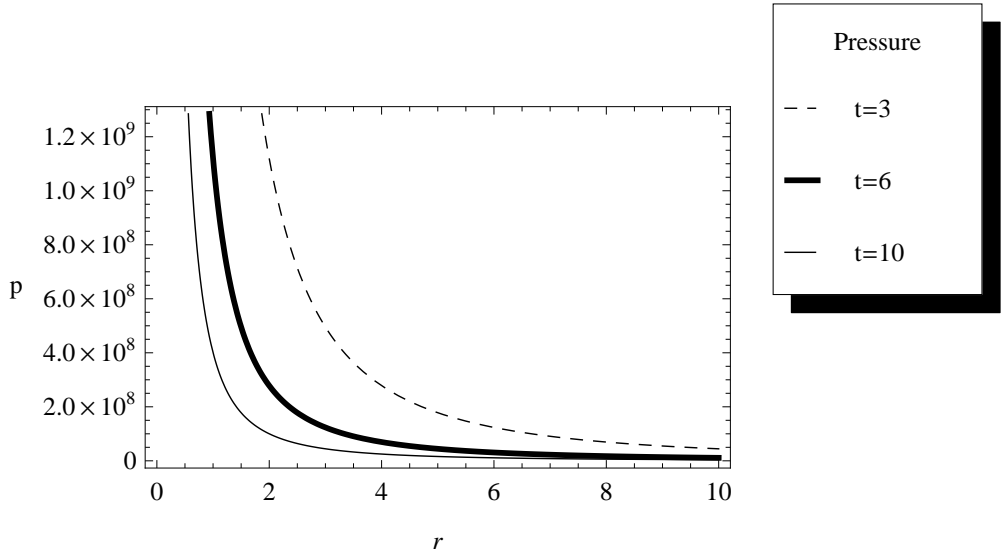} \caption{The radial profile for the pressure $p$ at various units of time.} \label{fig6}
\end{figure}

\begin{figure}[t]
\centering
\includegraphics[scale=1.5]{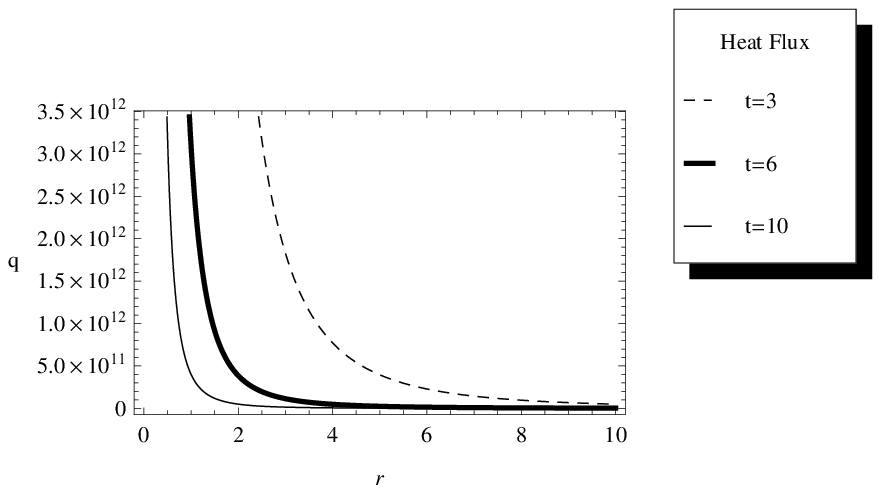} \caption{The radial profile for the heat flux $q$ at various units of time.} \label{fig6}
\end{figure}

\newpage
\begin{center}
\large{\bf Conflict of Interests}
\end{center}
The authors declare that there is no conflict of interests in regards to the publication of this paper.

\begin{center}
\large{\bf Acknowledgements}
\end{center}
B. P. Brassel, S. D. Maharaj and G. Govender thank the National Research Foundation and the University of KwaZulu-Natal for their financial support. S. D. Maharaj further acknowledges that this research is supported by the South African Research Chair Initiative of the Department of Science and Technology.

\end{document}